\input harvmac.tex
\noblackbox

\Title{CLNS 96/1415, USC-96-014}
{\vbox{
\centerline{Exact Friedel oscillations in the $g={1\over 2}$
Luttinger liquid. }
}}
\centerline{A. Leclair$^1$, F. Lesage$^2$, H.
Saleur$^2$}
\smallskip
\centerline{$^1$Newman Laboratory, Cornell University}
\centerline{Ithaca, NY 14853}
\smallskip
\centerline{$^2$Department of Physics}
\centerline{University of Southern California}
\centerline{Los Angeles, CA 90089-0484}

\vskip .3in

Abstract: A single impurity in  the 1D Luttinger model creates a
local  modification
of the charge density analogous to the Friedel oscillations. In this
paper, we present an exact
solution
of the case $g={1\over 2}$ (the equivalent of the Toulouse point) at
any
temperature $T$ and impurity coupling, expressing the charge density
in
terms of a hypergeometric function. We find in particular that at
$T=0$,
the oscillatory part of the density goes as $\ln x$ at small
distance
and $x^{-1/2}$ at large distance.

\Date{06/96}

\newsec{Introduction}

The Luttinger model, describing the low-energy excitations of an
interacting one-dimensional fermion gas, is one of the simplest
non-Fermi-liquid metals.
Experimental observations of this non-Fermi state in 1D quantum wires
are difficult, since disorder tends to localize the excitations. The
model
has also been proposed to describe the edge states in fractional
quantum Hall devices \ref\Wen{X. G. Wen, Phys. Rev. B41 (1990)
12838.}. Tunneling through a point contact is then a practically
ideal situation for comparing theory with experiments \ref\KF{C. L.
Kane, M. P. A. Fisher, Phys. Rev. B46 (1992) 25233.},
\ref\Metal{K. Moon, H. Yi, C. L. Kane, S. M. Girvin and M. P. A.
Fisher,
Phys. Rev. Lett. 71 (1993) 4381.}, \ref\FLS{P. Fendley, A. Ludwig, H.
Saleur,
Phys. Rev. Lett. 74 (1995) 3005.}.

While most of the attention has focussed on transport properties so
far,
static properties are also of interest. In particular,  the effect
of the impurity on the charge density, the equivalent of Friedel
oscillations \ref\Friedel{J. Friedel, Nuovo Cim. Suppl. 7 (1958)
287.},  has recently been considered \ref\Reinhold{R. Egger, H.
Grabert,
Phys. Rev. Lett. 75  (1995) 3505; R. Egger, H. Grabert, ``Friedel
oscillations in Luttinger
liquids'', in {\sl Quantum Transport in Semiconductor Submicron
Structures},
edited by B.Kramer, NATO-ASI Series E (Kluwer, Dordrecht, 1996)}.
A very similar problem, in antiferromagnetic spin chains,
was also discussed in \ref\affspin{I. Affleck, S. Eggert, Phys. Rev.
B46, 10866 (1992); Phys. Rev. Lett. 75, 934 (1995).}. See also
\ref\SchEck{P. Schmitteckert, U. Eckern, cond-mat/9604005.} for other
related
work.

The 1D
Luttinger model with an impurity is integrable, and ultimately, using
advanced  techniques
of quantum field theory \lref\Smir{F. Smirnov, {\sl Form factors in
completely
integrable models of quantum field theory}, World Scientific, and
references therein.}, \ref\LSS{F. Lesage, H. Saleur, S. Skorik,
Phys. Rev. Lett., vol. 76, (1996) 3388, cond-mat/9512087;
to appear Nucl. Phys. B (1996), cond-mat/9603043.},
these Friedel oscillations should be exactly computable for
all couplings, but
there are important  technical difficulties.  The particular point
$g=1/2$, the equivalent of the ``Toulouse point'' in the anisotropic
Kondo problem, is equivalent to free fermions \ref\GHM{F. Guinea, V.
Hakim, A. Muramatsu, Phys. Rev. Lett. 54 (1985) 263.}. It should thus
be  amenable by more elementary techniques. However, even in that
case,
exact expressions have not yet been obtained \Reinhold\  because the
density operator is not local in the Fermion basis. We show in this
paper how to circumvent this difficulty, largely based on a work
of Chatterjee and Zamolodchikov \ref\CZ{R. Chatterjee, A.
Zamolodchikov, Mod. Phys. Lett. A, vol. 9, (1994) 2227.}.

\newsec{Generalities}

Let us start from a description of
the quantities involved in the Luttinger liquid, following closely
\Reinhold.   Using
standard bosonization formula \ref\bos{J.M. Luttinger, J. Math. Phys.
4 (1963) 1154;
F.D.M. Haldane,  J. Phys. C14 (1981) 2585; Phys. Rev. B9 (1975)
2911.}, one can write the
electron creation
operator in the spinless case -  to which we restrict here -
as a
combination of two bosonic fields, $\phi(x), \theta(x)$. Decomposing
into left and right moving parts, one has :
\eqn\elecop{
\psi(x)^\dagger (\hbox{ resp. }\bar{\psi}(x)^\dagger)
\propto  \exp\left[ \pm i k_f x+
i\sqrt{\pi}(\theta(x)\pm \phi(x))\right].
}
The canonical momentum for $\phi$ is $\partial_x\theta$ and
they obey the commutation relations~:
\eqn\commrel{
[\theta(x),\phi(x')]=-{i\over 2} sgn (x-x').
}
The Friedel oscillations is describing the charge or density
oscillations
of the electrons in the presence of a barrier.  The density operator
  \ref\schulz{H.J. Schulz, Phys. Rev. Lett. 71 (1993)
1864.}~:
\eqn\densop{
\rho(x)=\rho_0+{1\over \sqrt{\pi}}\partial_x \phi +{k_f\over \pi}
 \cos[2k_f x+
2\sqrt{\pi} \phi(x)]
}
with $\rho_0={k_f\over \pi}$ the background charge.  The hamiltonian
for
these fields is~:
\eqn\halm{
H={v_f\over 2}\int_{-\infty}^\infty  dx \
[\Pi^2+(\partial_x\phi)^2]+H_{int}.
}
  The term $H_{int}$ describes a screened Coulomb
interaction.
 The impurity will be coupled to the fields
at one point $x=0$ by the term~:
\eqn\impo{
H_{imp}={\sqrt{\pi}\lambda\over k_F}\partial_x\phi(0)+\lambda
\cos[2\sqrt{\pi}\phi(0)].}
We will restrict to the  case  where $H_{int}$ is short range,
leading to a Luttinger liquid.
  In that case,
the effect of the interaction is to renormalize the fields.
The hamiltonian can then be brought into the usual form (setting
$v_F=g$)~:
\eqn\sing{
H=\int_{-\infty}^\infty  dx \
\left[8\pi g \Pi^2+{1\over 8\pi g} (\partial_x\phi)^2\right]+\lambda
\cos[\phi(0)],
}
while the Friedel oscillations, subtracting the background charge
density,
are encoded into~:
\eqn\fried{{<\rho(x)-\rho_0>\over\rho_0}=\cos \left(2k_F
x+\eta_F\right)
<\cos\phi(x)>.}
Here, the additional phase shift $\eta_F=-{g\pi\lambda \over k_F}$
in the cosine term arises from
the unitary transformation that cancels the $\partial_x\phi(0)$ term
in \impo\ to get \sing.

To  proceed, we perform
some manipulations.
Decompose the field $\phi$ into its left and right components,
$\phi=\phi_L+\phi_R$.
Introduce then the left movers~:
$$
\eqalign{\phi^e(x+t)={1\over\sqrt{2}}
\left[\phi_L(x,t)+\phi_R(-x,t)\right]\cr
\phi^o(x+t)={1\over\sqrt{2}}\left[\phi_L(x,t)-\phi_R(-x,t)\right]\cr}
$$
Now the impurity interaction reads
$H_{imp}=\lambda\cos[\sqrt{2}\phi^e(x=0)]$,
while the observable we are studying is proportional to~:
$$
\cos\phi(x,t)=\cos\left\{{1\over\sqrt{2}}
\left[\phi^e+\phi^o\right](x+t)
+{1\over\sqrt{2}}\left[\phi^e-\phi^o\right](-x+t)\right\}
$$
Set now~:
$$
\eqalign{&\Phi^e_L=\sqrt{2}\phi^e(x+t), \ x<0\quad
\Phi^e_R=\sqrt{2}\phi^e(-x+t), \ x<0\cr
&\Phi^o_L=\sqrt{2}\phi^o(x+t), \ x<0\quad
\Phi^o_R=-\sqrt{2}\phi^o(-x+t), \ x<0\cr}
$$
We then fold the system and recombine these left and right components
into a single field to get the hamiltonian~:
\eqn\newham{H=H^o+H^e,}
with~:
\eqn\newhami{H^o={1\over 2}\int_{-\infty}^0  dx \
\left[8\pi g (\Pi^o)^2+{1\over 8\pi g}
(\partial_x\Phi^o)^2\right],\quad
\Phi^o(0)=0,}
and~:
\eqn\newhamii{H^e={1\over 2}\int_{-\infty}^0  dx \
\left[8\pi g (\Pi^e)^2+{1\over 8\pi g} (\partial_x\Phi^e)^2\right]+
\lambda\cos{1\over 2}\Phi^e(0).}
We will often refer to \newhamii\ as the boundary sine-Gordon
hamiltonian.  Using parity together with the decoupling of
the odd and even fields we have~:
\eqn\newfried{{<\rho(x)-\rho_0>\over\rho_0}=\cos( 2k_Fx+\eta_F)
<\cos{1\over 2}\Phi^o(x)>
<\cos{1\over 2}\Phi^e(x)>.}

We shall obtain results in the continuum limit. Then, since
there is no boundary coupling for the odd field, one has
\ref\gogo{M. Fabrizio, A. Gogolin, Phys. Rev. B {\bf 51},
17827 (1995).}~:
\eqn\oddav{<\cos{1\over 2}\Phi^o(x)>\propto \left({a\over
x}\right)^{g/2},}
where $a$ is a lattice coupling ,  $x>>a$ ($a$ is defined
by the propagator $<\Phi^o(x)\Phi^o(0)>=-4g\ln (x/a)$ for the
bulk theory. The exact proportionality factor  in \oddav\  will
be worked out later).  Regarding the even field,
things are more complicated. On general grounds, one expects~:
\eqn\evnav{<\cos{1\over 2}\Phi^e(x)>\propto \left({a\over
x}\right)^{g/2}
F\left[\lambda a\left({x\over a}\right)^{1-g}\right],}
where the function $F$ goes to one for large values
of the argument,
and vanishes linearly at small values of the argument (this latter
property follows from the perturbative analysis in the regime
$g>1/2$).
Continuum limit results will hold for $x>>a$ and thus  $\lambda
<<1/a$, while the
product $\lambda a\left({x\over a}\right)^{1-g}$ is kept constant.
For simplicity we
set $a=1$  in what follows.

\newsec{Exact Friedel oscillations  at $g=1/2$ and $T=0$.}

The point $g=1/2$ corresponds to free fermions in the bulk, the
analog
of the Toulouse limit in the Kondo problem. To compute
the universal function $F$ exactly
in that case, we will proceed in two steps:
 (i) We will show that  $\cos{\Phi^e\over 2}$  can be expressed
as the product of spin operators in two decoupled
massless Ising models, one of them having a  boundary
magnetic field, the other having fixed boundary conditions.
(ii) These spin correlators in the boundary Ising models will be
computed based on a method due to   Chatterjee and Zamolodchikov.

In the following we do computations for
the even field (ie the field having boundary interaction);
results for the odd field follow
simply by taking the limit of Dirichlet boundary conditions
in the former solution.

\def\psibar{\overline{\psi}}
\def\psib{\psibar}
\def\gt{\tilde{\lambda}}

A natural way to relate the boundary sine-Gordon model to two copies
of the Ising model is by using bosonization.
At the free fermion point this was
considered in \ref\rkonlec{A. Ameduri, R. Konik and A. LeClair,
Phys. Lett. B 354 (1995) 376.}, where it was shown that the
boundary interaction $\lambda\cos{1\over 2}\Phi^e(0)$
becomes  linear in the complex Dirac fermion fields,
$\psi_\pm, \psibar_\pm$.  (In the conformal limit, $\psi_\pm$
and $\psibar_\pm$ are the left and right components and $\pm $
is the $U(1)$ charge.)  The boundary equations of motion at $x=0$
take the form \rkonlec:
\eqn\eIIxii{\eqalign{
\psi_+ +  \psi_-
-  \psib_- -  \psib_+ &= 0 \cr
i\partial_t ( \psi_- -  \psib_+ ) -  \gt^2
( \psib_-  - \psi_- ) &= 0 \cr
i\partial_t ( \psi_+ -  \psib_- ) +  \gt^2
( \psi_+  -  \psib_+ ) &= 0  ,  \cr}}
where $\gt \propto \lambda  $ and  $\lambda$ is
defined in \newhamii.
Note that these equations are valid regardless of whether there
is a bulk mass term or not.

Defining real components of the fermi fields as
$\psi_\pm = \psi_1 \pm i \psi_2$, $\psib_\pm =
- (\psib_1 \pm i \psib_2 )$, one finds that \eIIxii\ is equivalent to
\eqn\eisingbc{\eqalign{
\psi_1 &= - \psib_1 \cr
i\partial_t (\psi_2 - \psib_2 ) + \gt^2 (\psi_2 + \psib_2 ) &=\ \  0.
\cr }}
One can now compare these equations
with the boundary Ising equations of motion \CZ,
one sees that
 at the free fermion point, the boundary sine-Gordon theory
is equivalent to two boundary Ising models, one with fixed boundary
conditions (equivalently a magnetic field $h=\infty$), the other with
a varying
magnetic field $h \propto \lambda$.  The precise relation between
$h$ and $\lambda$ is however more delicate to obtain, see below.

A less natural, but more precise
 way to relate boundary sine-Gordon model
 to two copies of the Ising model is to use a scattering
description,
based on the boundary S-matrix and form factors.
To clarify some
of the following discussion, let us for a while
add to $H^e$  a term proportional to
$\int_{-\infty}^0 \cos\Phi^{e}(x)$, generalizing it into
a massive  boundary sine-Gordon model, which is well known
to be integrable \ref\GZ{S. Ghoshal, A. Zamolodchikov, Int. J. Mod.
Phys. A9 (1994) 3841.}.
 At $g=1/2$, to which we restrict, the spectra of the theory
is composed of solitons and anti-solitons (there is no
breathers) and the boundary reflexion
matrix is described by the two terms,
$P$ and $Q$  given by \GZ~:
\eqn\bmat{
R_\pm^\pm=P , \ \ R_\pm^\mp=Q ,
}
where the labels $\pm$ refer to  the soliton and
antisoliton of the sine-Gordon spectrum.
For our purposes it is much more convenient to take the
following combinations~:
\eqn\combin{\eqalign{
P+Q&=i\tanh\left(i{\pi\over 4}-{\theta\over 2}\right),\cr
  P-Q&=i\tanh\left(i{\pi\over 4}-{\theta\over
2}\right){\kappa-i\sinh\theta\over
\kappa+i\sinh \theta}.
}}
The key remark here is that
the combination given are just reflection matrices for the
fermion of the Ising model in a low temperature phase in
a boundary magnetic field $h$, with $\kappa = 1 - {4\pi h^2\over m}$.
($\kappa={1\over k}$ in notations of \GZ).
One combination  describes fixed Ising spin boundary
conditions (or infinite boundary magnetic field), and the
other is at finite Ising  boundary magnetic field.  This suggests
that in
  the basis of symmetric and antisymmetric
combinations of  solitons and anti-solitons,
the theory decouples to two Ising models with boundary magnetic
fields.

This can be confirmed for instance by considering  form factors, ie
matrix elements of the operators in the quasiparticle basis \Smir.
The form factors for the operators $e^{\pm i\Phi/2}$ in the
sine-Gordon model at $g=1/2$
are given by~:
\eqn\ffavecb{
f^{\pm}(\theta_1,...\theta_{2n})_{\epsilon_1,...,\epsilon_{2n}}=
 (\pm 2i)^n
\sqrt{u_1...u_{2n}} \prod_{i=1}^{2n} u_i^{\mp \epsilon_i/2}
\prod_{i<j} (\epsilon_i u_i-\epsilon_ju_j)^{\epsilon_i \epsilon_j}
}
with the notation $u_i=e^{\theta_i}$, $\theta$ the usual quasi
particle
rapidity \Smir.  In order to show the relation
to the Ising model, we change the basis from
$|A>$, $|S>$ to ${1\over \sqrt{2}}(|S>+|A>)$ and ${1\over
\sqrt{2}}(|S>-|A>)$.
The explicit calculation is done in
the appendix, in the new basis the form factors are~:
\eqn\ina{
f^\pm_{\epsilon_1',...,\epsilon_{2n}'}(\theta_1,...,\theta_{2n})=
 i^n(-1)^{n_-/2}
\prod_{i<j} \tanh\left( {\theta_i-\theta_j\over 2}\right)^{\vert
\epsilon_i'+\epsilon_j'\vert/2},
}
where the isotopic indices refer now to the sign of the
combination between solitons and anti-solitons and
$n_-$ denotes the number of $\epsilon_i'$ indices having
value $-$. This formula holds only for $n_-$ even, which
will be the case for the correlator of interest.   Recall now
the expression of form factors for spin operator $\sigma$
in the Ising model \ref\isingff{B. Berg, M. Karowski, P. Weisz,
Phys.s Rev. D19 (1979) 2477; V. P. Yurov, Al. B. Zamolodchikov, Int.
J. Mod. Phys. A6 (1991) 3419; J. L. Cardy, G. Mussardo, Nucl. Phys.
B340 (1990) 387.}~:
\eqn\isingff{f^{\sigma}(\theta_1,\ldots,\theta_{2n})
=i^{n}\prod_{i<j}\tanh {\theta_i-\theta_j\over 2},}
the form factors of $\mu$ on an even
number of particles vanishing.

In the presence of the boundary, the one point function of
of interest  is obtained  as $<B|\cos {\Phi^e(x)\over 2}|0>$, where
$|B>$
is the boundary state, $|0>$ the ground state. The
boundary state can be expressed in terms
of multiparticle states following \GZ. In the new basis
mixing solitons and anti-solitons, this boundary state
actually factorizes~:
\eqn\bstate{\eqalign{
|B>&=|B>_+\otimes |B>_-,
\cr
|B>_{\pm}&=\exp\left[ \int {d\beta\over 2\pi} A_\pm^*(-\beta)
A_\pm^*(\beta) K_{\pm }(\beta)
\right]
}}
where $A^*_{\epsilon'}$   denote the creation operators in the new
basis,
and $K_\pm(\theta)\equiv \pm(P\pm Q)({i\pi\over 2}-\theta)$. Here,
the additional $\pm$ sign occurs because the $K$ matrix is obtained
through the\
general formula \GZ\ $K^{ab}(\theta)=R^b_{\bar{a}}\left({i\pi\over
2}-\theta\right)$; soliton and antisoliton are conjugate in the
sine-Gordon model, while the Ising fermion is self conjugate.
Expanding these
boundary
states, using the form factors \ina\ and \isingff, we see that the
correlation function becomes indeed  a product
of two spin operators (in the low temperature phase of the
Ising models, where the one point function of the
disorder operator vanishes)~:
\eqn\prdis{
<\cos {\Phi^e\over 2}>=<\sigma>_h <\sigma>_\infty.
}

Having shown  \prdis, we may now let  the mass of the sine-Gordon
model
(the
amplitude of the bulk $\cos\Phi^e$ term) go to zero to recover the
original
problem.
The correspondence between the variable $\lambda$
in the boundary sine-Gordon, and the field $h$ in the
boundary Ising model, can be found in that case using results
of \ref\FLS{P. Fendley, A. Ludwig, H. Saleur, Phys. Rev. B., vol. 52,
(1995) 8934.}.
Using
the conventions of \CZ\ for the boundary Ising model
one finds $T_B=8\pi h^2=4\pi \lambda^2$ so
$h={\lambda\over\sqrt{2}}$.  This completes our derivation
of the relation between the Friedel oscillations and the
Ising model.   We can now  concentrate on finding the correlation
of the spin  operator in the Ising model  with
a boundary magnetic field.

At $T=0$, the one point function of the spin with a boundary
has been evaluated in the very
interesting paper \CZ.
There it is shown
that (we trade the variable $\lambda$ of the original action for
$h$), introducing the
variable $X=4\pi h^2 x$, and setting
$\bar{\sigma}(X)=\left<\sigma(x)\right>_h$, the following holds~:
\eqn\diffeq{\left[ 4{d^2\over d X^2}+
\left({1\over X}-8\right){d\over dX}
+\left(-{1\over X}+{9\over 16}{1\over
X^2}\right)\right]\bar{\sigma}(X)=0.}
The solution of this equation that describes
the appropriate physics is~:
\eqn\soleqd{\left<\sigma(x)\right>_h=2^{13/8}\sqrt{\pi}h x^{3/8}
\Psi(1/2,1;8\pi h^2 x),}
where $\sigma$ is normalized as usual in the bulk.
Here, $\Psi$ is a degenerate hypergeometric function. It is simply
expressed in terms of Bessel functions as
$\Psi(1/2,1;2x)=
{e^x\over \sqrt{\pi}}K_0(x)$.
 At short distance one finds that $\sigma\approx -2^{13/8}h\
x^{3/8}\ln x$
and at large distances $\sigma\approx 2^{1/8}x^{-1/8}$.
Taking
into account the fact that the physical observable in the
case of Friedel oscillations is \newfried\ one finds~:
\eqn\newfriedfinal{{<\rho(x)-\rho_0>\over\rho_0}=4\sqrt{\pi} h\
\cos( 2k_Fx +\eta_F)
\Psi(1/2,1;8\pi h^2 x), \ \ h={\lambda\over\sqrt{2}}.}
The asymptotic behaviour is ${<\rho(x)-\rho_0>\over\rho_0}\propto
-\ln x\ \cos (2k_F x+\eta_F)$ as $x\to 0$ and
${<\rho(x)-\rho_0>\over\rho_0}\propto x^{-1/2}\ \cos (2k_F x+\eta_F)
$
as $x\to\infty$.

Our result agrees with the numerical simulations in \Reinhold, except
at very small $x$, where it was found that
${<\rho(x)-\rho_0>\over\rho_0}$
behaves as a power law, with a small negative exponent. This
discrepancy
is very likely due to fact that, for  stability reasons, the true UV
region
is difficult to access numerically - indeed, at intermediate values
of $x$, our
result does behave like a power law \ref\reinholdlast{R. Egger,
private communication.}.

\newsec{Exact Friedel oscillations  at $g=1/2$ and $T\neq0$.}

We can also extend  the  computation of \CZ\
to finite temperature. The main idea
is still  that
the boundary magnetic field does not destroy the free
field structure of the Ising model \ref\McCWu{B. Mc. Coy, T. T.  Wu,
{\sl The Two Dimensional Ising Model}, Harvard University Press
(1973)}, leading
to a determination of the correlator
by elementary considerations \CZ.

For convenience, we first rotate the geometry
so now the boundary lies along the $x$ axis.
As mentioned earlier \eIIxii,
the equations of motion for the
Majorana fermion of the Ising model
with boundary magnetic field are (in imaginary time)
{}~:
\eqn\bcond{
[(\partial_z+4i\pi h^2)\psi(z)-(\partial_{\bar{z}}-4i\pi
h^2)\bar{\psi}
(\bar{z})]_{z=\bar{z}}=0,
}
where $z=x+iy$. Having this condition, the idea is to
introduce a fermion field $\chi(z)=(\partial_z+4i\pi h^2)\psi(z)$
and its conjugate $\bar{\chi}(\bar{z})$, and realize that the
previous boundary condition
states that
$\bar{\chi}(\bar{z})$
is the analytic continuation of $\chi(z)$ to the lower half plane.
Hence the correlator $<\chi(z)\mu(w,\bar{w})>$
is analytic in the {\bf full} $z$ plane
with two square root branch points at $z=w$ and $z=\bar{w}$.

With a finite temperature, the argument
is the same apart from the periodicity
in the imaginary time direction, which
after having rotated the system is the $x$ direction, is $z\to
z+{1\over\beta}$.  The boundary condition, a local statement,
remains the same.

The next step is to write a global form for the correlator
$<\chi(z)\mu(w,\bar{w})>$, and this is where the effect of
temperature will be seen.  On the cylinder,
one requires the right hand side to have
square root branch points at $w+n T,\bar{w}+mT$,
$n,m$ integers, and to be periodic.  One can therefore write~:
\eqn\equivform{\eqalign{<\chi(z)&\mu(w,\bar{w})>\left({\sin \pi
T(z-w)\over\pi T}
\right)^{1/2}
\left({\sin \pi T (z-\bar{w})\over\pi T}\right)^{1/2}\cr
&=\pi T\ \cot[\pi T(z-w)]\ A(w,\bar{w},T)
+\pi T\ \cot[\pi T(z-\bar{w})]\
\bar{A}(w,\bar{w},T)+B(w,\bar{w},T).\cr}}
Observe how the right hand side
is periodic in $z\to z+ T$.
In this form, the coefficients, $A, \bar{A}, B$ are unknown that
need to be fixed.
We can now use the operator product
expansions, for example in $(z-w)$ ($\omega=e^{i\pi/4}$)~:
\eqn\operprod{
\eqalign{\chi(z)\mu(w,\bar{w})=
{\bar{\omega}\over\sqrt{2}}(z-w)^{-3/2}
\{-{1\over 2}\sigma(w,\bar{w})+(z-w)
(2\partial_w+4i\pi h^2)\sigma(w,\bar{w})\cr
+(z-w)^2(4\partial^2_w+16i\pi h^2\partial_w)
\sigma(w,\bar{w})+\ldots\} .\cr}}
and a similar relation for $\bar{\chi}$.
The reader might fear that
\operprod\
which holds in the plane might be changed by $\beta$ dependent terms
when the geometry is compactified. Actually, it is well known
\ref\BPZ{A.A. Belavin,
A.M. Polyakov, A.B. Zamolodchikov, Nucl. Phys. B 241, (1984) 333.}
that short distance expansions are invariant in conformal mappings.
This
constraint for instance determines the
coefficients $4,8/3$ in the operator product \CZ\ :
\eqn\opepsimu{\psi(z)\mu(w,\bar{w})={\bar{\omega}\over\sqrt{2}}
(z-w)^{-1/2}\left[\sigma(w,\bar{w})+4(z-w)\partial_w\sigma(w,\bar{w})
+{8\over 3}(z-w)^2\partial_w^2\sigma(w,\bar{w})+\ldots\right].}
Now taking the global form \equivform\ and expanding in
$(z-w)$ we get relations between the coefficients $A,\bar{A},B$ and
the spin function and its derivatives (similarly expanding in
$(\bar{z}-\bar{w})$ and comparing with the OPE of $\bar{\chi}\mu$
we get more relations).  Then,
by matching \equivform\ and \operprod\ and eliminating the
unknown coefficients one finds that the
one point function of the spin satisfies the equation~:
\eqn\monster{\eqalign{&\left\{{d^2\over (d y)^2}
+\left({\pi T\over 2}\coth 2\pi
Ty-8\pi h^2\right){d\over d y}\right.\cr &
\left.+\left[{9\over 16}(\pi T)^2(\coth 2\pi Ty)^2-
{4\pi^2h^2 T\over 2}
\coth 2\pi Ty-{\pi^2T^2\over 2}\right]\right\}<\sigma(y)>=0.\cr}}
In particular, this equation reproduces eq. (30) of \CZ\ in the limit
$T\to 0$.
To solve this equation, let us set
$<\sigma(y)>=\left[\sinh (2\pi Ty)\right]^{-1/8} f(y)$. Introduce
then the
new variables $\Lambda={2h^2\over  T}$ and $Y=(1-\coth 2\pi Ty)/2$,
and
set $f(y)=\bar{f}(Y)$.  We then find
an hypergeometric equation~:
\eqn\monsterii{\left\{(Y-Y^2){d^2\over(d Y)^2}
+\left(1+{\Lambda\over 2}-2Y\right){d \over d Y}-{1\over 4}\right\}
\bar{f}(Y)=0.}
The physical solution of the problem , going back
to the original $x$ variable, is then~:
\eqn\nice{<\sigma(x)>_h=\left({4\pi T\over \sinh 2\pi
Tx}\right)^{1/8}
F\left({1\over 2},{1\over 2};1+2{h
^2\over  T},{1-\coth 2\pi Tx\over 2}\right)
 .}
To select this solution, we observe that the one
point function at large $x$ can should expand, by conformal
invariance, as a sum of exponential terms $\exp(-4\pi Tx\Delta)$,
with
$\Delta$ conformal weights of the  (central charge) $c={1\over 2}$
conformal
field theory. This excludes
the other  hypergeometric function solution of \monsterii.   The
normalisations are fixed by looking at the case $h\rightarrow
\infty$.
If $h\to\infty$, one has fixed boundary conditions
for the spin. The one point function can then be obtained
by conformal mapping from the half plane where it is known
to be \ref\CardyLew{J. Cardy, D. Lewellen, Phys. Lett. 259B (1991)
274.}
$<\sigma(x)>_{hp}=2^{1/8}x^{-1/8}$ (with
$<\sigma(x)\sigma(0)>=x^{-1/4}$ in the full
plane). Setting $z'= {1\over 2\pi T}\ln
{z-1\over z+1}$
maps the half plane on the semi-infinite cylinder, the line  boundary
$Re z=0$
becoming the circle boundary $Re z'=0, Im z' \in [-1/2T,1/2T]$. In
this
transformation one finds for the semi-infinite cylinder
$<\sigma(x)>_{sic}=
\left({4\pi T\over \sinh 2\pi Tx}\right)^{1/8}$, in agreement
with \nice. Also, as $x\to\infty$, we expect to
recover the same result since at large  distance the
spin sees fixed boundary conditions. This fixes the normalization
constant
in \nice.

Finally, we can study the limit $T\to 0$. By using the standard
transformation formulas $z\to 1/z$ for the argument of the
hypergeometric
functions, together with the definition of $\Psi$ in terms
of the basic degenerate hypergeometric function $\Phi$ (all notations
are those
of \ref\GR{I. S. Gradshteyn, I. M. Ryzhik, {\sl Table of Integrals,
Series and Products}, Academic Press.}), one
finds, as $\gamma,z\to\infty$, $\gamma/z$ finite
$$
F(\alpha,\alpha;\gamma,z)\approx\left({-\gamma\over z}\right)^\alpha
\Psi(1/2,1;-\gamma/z),
$$
where we used
$$
F(\alpha,\alpha+1-\gamma;1,1/z)\approx \Phi(\alpha,1;-\gamma/z),
$$
which can be proved using the series representation of the involved
functions. Therefore, as $T\to 0$, one finds $<\sigma(x)>_h\approx
2^{1/8}
x^{-1/8} (8\pi h^2 x)^{1/2}\Psi(1/2,1;8\pi h^2 x)$, in agreement
with the  foregoing results \CZ.

The qualitative effect of the temperature can be seen
on the small and large $x$ limits of the one point function. One has
\eqn\limtt{\eqalign{<\sigma(x)>_h&\approx
-2^{9/8}T^{1/2}{\Gamma\left(1+{2h^2\over T}\right)
\over \Gamma\left({1\over 2}+{2h^2\over T}\right)},\ \  x<<1/T\cr
<\sigma(x)>_h&\approx (8\pi)^{1/8} e^{-\pi Tx/4},\ \  x>>1/T.\cr}}

We can now come back to our original problem, the Friedel
oscillations.  Using \newfried\ with the previous solution we
find the final expression~:
\eqn\final{
{<\rho(x)-\rho_0>\over \rho_0}=\cos (2k_F x+\eta_F)
\left( {4\pi T\over \sinh 2\pi T x}\right)^{1/2} F\left( {1\over 2},
{1\over 2}; 1+2{h^2\over T}, {1-\coth 2\pi T x\over 2}\right),
}
where as before $h={\lambda\over \sqrt{2}}$.
The periodicity $x \to x -i/T$ of the final solution might
appear a bit surprising since  it is the imaginary time $y$
which is compactified on a circle of radius $1/T$, not $x$.
It
can be understood  from the fact that
one is dealing with a  {\it massless} theory in the bulk.
In fact, this periodicity can also be seen by taking the massless
limit of the expressions for finite temperature correlators obtained
in \ref\rllss{A. Leclair, F. Lesage, S. Sachdev and H. Saleur,
``Finite Temperature Correlations in the One-Dimensional
Quantum Ising Model'', cond-mat/9606104.} by using the form factors.

We finally
observe that the same technique could be applied to study the
screening
cloud at the Toulouse point of the anisotropic Kondo problem
\ref\Aff{I. Affleck, E. Sorensen, Phys. Rev. B53, 9153 (1996),
cond-mat/9511031}. This will be described elsewhere.

\bigskip
{\bf Acknowledgements:} We thank R. Egger for pointing out
this problem to us,  for  many useful
conversations, and for kindly checking our formula
against his numerical results.

\appendix{A}{Change of basis for the sine-Gordon form factors.}

In this appendix we want to show how the one particle form
factor for the operator $e^{\pm i\phi/2}$ in the sine-Gordon
model at the free fermion point is related to spin form-factor
in the  Ising model.  The form factor is a tensor function in the
space of
isotopic indices and in the new basis described by
${1\over\sqrt{2}}(|S>+|A>)$ and ${1\over\sqrt{2}}(|S>-|A>)$ (which we
will denote by $+$ and
$-$ in the following), the form factor is simply given by~:
\eqn\bastrf{
f_{\epsilon_1',...,\epsilon_{2n}'}(\theta_1,...,\theta_{2n})=
A^{\epsilon_1}_{\epsilon_1'}\cdots A^{\epsilon_{2n}}_{\epsilon_{2n}'}
f_{\epsilon_1,...,\epsilon_{2n}}(\theta_1,...,\theta_{2n}),
}
where the primed indices denote the new basis and
with~:
\eqn\tramat{
A^\epsilon_{\epsilon'}={1\over\sqrt{2}}\pmatrix{1&1\cr 1&-1}
}
describing the change of basis.  The forms factor in the $|S>, |A>$
basis was given before~:
\eqn\ffsin{
f^{\pm}(\theta_1,...\theta_{2n})_{\epsilon_1,...,\epsilon_{2n}}=
 (\pm 2i)^n
\sqrt{u_1...u_{2n}} \prod_{i=1}^{2n} u_i^{\mp \epsilon_i/2}
\prod_{i<j} (\epsilon_i u_i-\epsilon_ju_j)^{\epsilon_i \epsilon_j}
}
where the non-zero form factors are those preserving charge and we
used the variables $u_i=e^{\theta_i}$.

Let us first look at the form factor with all primed indices chosen
to
be $+$, it is given by~:
\eqn\plup{\eqalign{
f^\pm_{+...+}(\theta_1,...,\theta_{2n})=&
 (\pm i)^n \cr &
\sum_{\sum\epsilon_i=0}
\sqrt{u_1...u_{2n}} \prod_{i=1}^{2n} u_i^{\mp \epsilon_i/2}
\prod_{i<j} (\epsilon_i u_i-\epsilon_ju_j)^{\epsilon_i \epsilon_j}.
}}
By making a dilatation one sees hat this is an homogeneous
function of degree zero.  This function has poles for all values of
$u_i=-u_j$
and zeroes for $u_i=u_j$.  Only the sign has to be worked out,
the final form factor is~:
\eqn\ffpp{
f^\pm_{+...+}(\theta_1,...,\theta_{2n})=
 i^n
\prod_{i<j} \tanh\left( {\theta_i-\theta_j\over 2}\right).
}
Imagine now that some of the primed indices are $-$, then the
function
is still homegeneous of degree zero but if let say $\phi_i$ are the
rapidities
of the particles having indices $-$, it is not difficult to show that
there are now no zeroes and poles between terms involving
$\theta_i$ and $\phi_j$.  There are still zeroes and poles between
each type of particle and we find that the form factor is given by~:
\eqn\fina{
f^\pm_{\epsilon_1',...,\epsilon_{2n}'}(\theta_1,...,\theta_{2n})=
  i^n (-1)^{ n_-/2}
\prod_{i<j} \tanh\left( {\theta_i-\theta_j\over 2}\right)^{\vert
\epsilon_i'+\epsilon_j'\vert/2}.
}
In this expresion, $n_-$ is the number of indices
$\epsilon_i'$ of type $-$ (\fina\ holds only when $n_-$ is even,
which is
the case for our computation).

\listrefs
\bye